\documentclass{llncs}
\usepackage{amsmath} 
\usepackage{textcomp} 
\usepackage{pstricks} 
\usepackage{pst-node} \usepackage{prooftree} 
\usepackage{coqdoc} 
\usepackage{url} 
\usepackage{xy}
\usepackage{qsymbols}

\newif\ifcomment
\commenttrue 
\newcount\timehh\newcount\timemm \timehh=\time
\divide\timehh by 60 \timemm=\time
\newcommand{\timestamp}{
  {\protect\small\sl\today\ --
    \ifnum\timehh<10 0\fi\number\timehh\,:\,
    \ifnum\timemm<10 0\fi\number\timemm}}

\newcommand{\Coq}{\textsc{Coq}} 
\newcommand{\ie}{i.e.,~} 
\newcommand{\leut}{\ensuremath{\preceq}} 

\newcommand{\hbis}{\sim_{h}}
\newcommand{\sbis}{\sim_{s}}
\newcommand{\lft}{\ensuremath{\ell}}
\newcommand{\ltl}{\textit{``leads to a leaf''}}
\newcommand{\altl}{\textit{``always leads to a leaf''}}

\newcommand{\coconva}{\ensuremath{\mathop{"<<-"\!
\raisebox{1.5pt}{\(\scriptstyle a\)}\! "->>"}}} 
\newcommand{\conva}{\ensuremath{\mathop{\vdash\! \raisebox{2pt}{\(\scriptstyle a\)}\! \dashv}}} 
\renewcommand{\H}{\ensuremath{\mathcal{H}}}
\newcommand{\og}{\ensuremath{\ll\!}}
\newcommand{\fg}{\ensuremath{\hspace*{-1.5pt}\gg}}

\title{Deconstruction of Infinite Extensive Games\\using Coinduction}

\author{Pierre Lescanne\thanks{This research has been supported by R\'{e}gion \emph{Ile de France}.}}  \institute{Universit\'e de Lyon, ENS de Lyon, CNRS (LIP), \\ 46 all\'ee
d'Italie, 69364 Lyon, France}

\begin{document} 
\newqsymbol{"|-|"}{\conva} 
\newqsymbol{"<<-a->>"}{\coconva}

\maketitle

\begin{abstract} 
\ifcomment \centerline{\timestamp} \fi

  Finite objects and more specifically finite games are formalized using induction, whereas infinite objects are formalized using \emph{coinduction}.  In this
  article, after an introduction to the concept of coinduction, we revisit on infinite (discrete) extensive games the basic notions of game theory.  Among others, we
  introduce a definition of \emph{Nash equilibrium} and a notion of \emph{subgame perfect equilibrium} for infinite games.  We use those concepts to analyze well known infinite
  games, like the \emph{dollar auction game} and the \emph{centipede game} and we show that human behaviors that are often considered as illogic are perfectly rational, if one admits
  that human agents reason coinductively.
\end{abstract}

\section{Introduction} 

Finite extensive games have been introduced by Harold Kuhn~\cite{Kuhn:ExtGamesInfo53}.  But many interesting extensive games are infinite and therefore the theory of infinite extensive games
play an important role in game theory, with examples like the \emph{dollar auction
  game}~\cite{Shubik:1971,colman99:_game_theor_and_its_applic,gintis00:_game_theor_evolv,osborne94:_cours_game_theory} or the generalized \emph{centipede game}.\footnote{Here
  \emph{``generalized''} means that the game has an infinite ``backbone''.}  But, from a formal point of view, they are not appropriately treated in papers and books. In particular,
there is no clear notion of Nash equilibrium and the gap between finiteness and infiniteness is not correctly understood.  Instances of vague definitions related to infiniteness
are in~\cite{osborne04a} p.~157 and in~\cite{osborne94:_cours_game_theory} p.~90, where the authors speak about the \emph{length of the longest history} without guaranteeing that
such a longest history exists.  More precisely, Osborne in~\cite{osborne04a} writes:
\begin{it}
    If the length of the longest derivation is [...] finite, we say that the game has a finite horizon.
    Even a game with a finite horizon may have infinitely many terminal histories, because some player has infinitively many actions after some history.
\end{it}
\noindent
If a game has only finite histories, but has infinitely many such finite histories of increasing length, the length of the longest history does not exist and it is not clear whether
such a game should be considered to have a \emph{finite horizon} or not.  

Another common mistake arises in escalation~\cite{Shubik:1971}. It is to believe that results on infinite games can be obtained as the limit of results on finite games.  It is
notorious that there is a threshold (a wall) between finiteness and infiniteness, a fact known to model theorists~\cite{fagin93:_finit_model_theor_person_persp,EF-finite-mt} and to
specialists of real variable functions, as shown by the Weierstrass function, a counterexample of the fact that a finite sum of functions differentiable everywhere is
differentiable everywhere whereas an infinite sum can be differentiable nowhere.  In this kind of inadequate reasoning, people study finite games that are the infinite games
truncated at a finite bound, then they extrapolate their results to infinite games by increasing the size to infinity.  But this says nothing since the \emph{infiniteness is not
  the limit of finiteness}.  They use this to conclude that humans are wrong in their reasoning.  But clearly, if there would be a limit, then there would be no escalation.  In
other words, if there is an escalation, then the game is infinite, then \emph{the reasoning must be specific to infinite games, \ie based on coinduction} and this is only on this
basis that one can conclude that humans are rational or irrational and for us it turns out that agents are rational in escalating.  In short, \emph{Macbeth is rational.}

In this paper we address these issues and the games we consider may have arbitrary long histories as well as infinite histories.  In our games there are two choices at each node,
this is without loss of generality, since we can simulate finitely branching games in this framework.  By K\"onig's lemma, finitely branching, specifically binary, infinite games
have at least an infinite history.  We are taking the problem of defining formally infinite games, infinite strategy profiles, and infinite histories extremely seriously.  Another
important issue which is not considered in the literature is how the utilities associated with an infinite history are computed.  To be formal and rigorous, we expect some kinds of
recursive definitions, more precisely co-recursive definitions, but then comes the questions of what the payoff associated with an infinite strategy is and whether such a payoff
exists.

Finite extensive games are represented by finite trees and are analyzed through induction.  For instance, in finite extensive games, a concept like \emph{subgame perfect
  equilibrium} is defined inductively and receives appropriately the name of \emph{backward induction}.  Similarly \emph{convertibility} (an agent changes choices in his strategy)
has also an inductive definition and this concept is a key for this of \emph{Nash equilibrium} which is not itself given by an inductive definition.  But \emph{induction}, which
has been designed for finitely based objects, no more works on infinite\footnote{In this paper, \emph{infinite} means infinite and discrete.  For us, an infinite extensive game is
  discrete and has infinitely many nodes.  } games, \ie games represented by infinite trees.  Logicians have proposed a tool, which they call \emph{coinduction}, to
reason on infinite objects.  In short, since objects are infinite and their construction cannot be analyzed, coinduction ``observes'' them, that is looks at how they ``react'' to
operations (see Section~\ref{sec:inf_obj} for more explanation).  In this paper, we formalize with coinduction, the concept of infinite game, of infinite strategy profile, of
equilibrium in infinite games, of utility (payoff), and of subgame and we verify on the proofs assistant \Coq{} that everything works smoothly and yields interesting consequences.
Thanks to coinduction, some examples of apparently paradoxical human behavior are explained logically, demonstrating a rational behavior.

\subsubsection{Induction vs coinduction.}
\label{sec:indvscoind}

To formalize structured finite objects, like finite games, one uses \emph{induction}, \ie a definition of basic objects (in the case of finite games they are leafs or terminal
nodes) and a definition of the way to build new objects (in the case of finite games induction provides an operation to build a game from subgames).  In the case of infinite objects like infinite
games, this no more works and one characterizes infinite objects by their behavior. But this raises two questions: what happens if I~query the utility (payoff) of an infinite
strategy profile?  What happens if I ask for a subgame of an infinite game?  This characterization by ``observation'' is called \emph{coinduction}.  Whereas induction is associated
with the least fixed point of a definition, coinduction is associated with the greatest fixed point.  The proof assistant \Coq{} offers a framework for coinductive
definitions and reasonings which are keys of our formalization.

 \subsubsection{Structure of the paper.}

 Deconstruction is a sketch of what coinduction does, but we can say that in this paper we undertake a deconstruction of the concept of infinite games using coinduction, which was
 made possible by the extreme rigor and discipline imposed by the proof assistant \Coq.  The proof assistant \Coq{} was indeed crucial for the author, but this paper tries to hide this aspect.

The paper is structured as follows. First we present the notion of infinite object using this of \emph{history} (Section~\ref{sec:inf_obj}), we use this example to introduce the
concept of coinduction.  Then we explore the notions of \emph{infinite game} (Section~\ref{sec:games}) and \emph{infinite strategy} (the word we use for \emph{strategy profile})
(Section~\ref{sec:inf_stra}).  The concept of \emph{convertibility} (Section~\ref{sec:conv}) allows us to introduce the concepts of \emph{equilibrium} (Section~\ref{sec:equilib}).
We apply those concepts to two infamous examples (Section~\ref{sec:inf_ped}), namely the \emph{dollar auction game} and the \emph{centipede game} and we say a few words about human
reasoning versus coinductive reasoning (Section~\ref{hum_form}). In Section~\ref{sec:coq} we tell how \emph{coinduction} is implemented in \Coq{} and how it is used.\footnote{This
  section can be dropped by readers less interested by the behavior of \Coq.} Related works are presented in Section~\ref{sec:rel_works}. Appendix gives formal version of the
definitions that are given in the paper, they are also for the reader interested by the formal aspects.

 \subsubsection{\Coq{} scripts.}

 The documentation of the \Coq{} scripts containing only statements of definitions and lemmas is given in
\begin{center}
  \url{http://perso.ens-lyon.fr/pierre.lescanne/COQ/INFGAMES/}
\end{center}
It contains what has been proved without the proofs themselves. The \Coq{} scripts, \ie the detail of the proofs, are in
\begin{center}
  \url{http://perso.ens-lyon.fr/pierre.lescanne/COQ/INFGAMES/SCRIPTS/}
\end{center}
This part is meant for those with some expertise in \Coq{} and who are interested by the correction of the logical arguments.

\section{An example of infinite objects:  histories}
\label{sec:inf_obj}

Infinite objects have peculiar behaviors. To start with a simple example, let us have a look at \emph{histories} in games. In a game, agents make \emph{choices}.  In an infinite
game, agents can make finitely many choices before ending, if they reach a terminal node, or infinitely many choices, if they run forever. Choices are recorded in a \emph{history}
in both cases.  A history is therefore a finite or an infinite list of choices.  In this paper, we consider that there are two possible choices: \lft~and \textsf{r} (\lft{} for
``left'' and \textsf{r} for ``right'').  Since a history is a potentially infinite object, it cannot be defined by structural induction.\footnote{In type theory, a type of objects
  defined by induction is called an \coqdockw{Inductive}, a shorthand for \emph{inductive type}.}  On the contrary, the type\footnote{Since we are in type theory, the basic concept
  is this of \emph{``type''}. Since we are using only a small part of type theory, it would not hurt to assimilate naive types with naive sets.} \emph{History} has to be defined as
a \coqdockw{CoInductive}, \ie by coinduction.  Let us use the symbol~$[~]$ for the empty history and the binary operator $::$ for non empty histories.  When we write $c::h$ we mean
the history that starts with $c$ and follows with the history $h$.  To define histories coinductively we say the following:
\begin{quotation}
  A \textbf{coinductive} history (or a finite or infinite history) is
  \begin{itemize}
  \item either the empty history $[\ ]$,
  \item or a history of the form $c::h$, where $c$ is a choice and $h$ is  a history.
  \end{itemize}
\end{quotation}
The word ``coinductive'' says that we are talking about finite or infinite objects.  This should not be mixed up with finite histories which will be defined inductively as follows:\pagebreak[3]
\begin{quotation}
  An \textbf{inductive} history (or a finite history) is built as
  \begin{itemize}
  \item either the empty history $[\ ]$,
  \item or a finite non empty history which is the composition of a choice $c$ with a finite history $\ell$ to make the finite history $c:: \ell$.
  \end{itemize}
\end{quotation}
Notice the use of the participial ``built'', since in the case of induction, we say how objects are built, because they are built finitely.    Let us now consider four families of histories:

\medskip

\begin{tabular}[h]{|l|l|}
  \hline
  $\H_0$ & {\scriptsize The family of finite histories}\\
  $\H_1$ & {\scriptsize The family of finite histories or of histories which ends with an infinite sequence of \lft's}\\
  $\H_2$ & {\scriptsize The family of finite histories or infinite histories which contains infinitely many \lft's}\\
  $\H_{\infty}$ & {\scriptsize The family of finite or infinite histories}\\
  \hline
\end{tabular}

\medskip

We notice that $\H_0\subset\H_1\subset\H_2\subset\H_{\infty}$.  If $\H$ is a set of histories, we write $c::\H$ the set $\{h \in \H_{\infty} \mid \exists h' \in \H, h = c :: h'\}$.
We notice that $\H_0$, $\H_1$, $\H_2$ and $\H_{\infty}$ are solutions of the fixpoint equation :
\[\H = \{[~]\} \ \cup\ \lft::\H \ \cup\ \textsf{r}::\H.\]
in other words
\begin{eqnarray*}
  \H_0 &=& \{[~]\} \ \cup\ \lft::\H_0 \ \cup\ \textsf{r}::\H_0\\
  \H_1 &=& \{[~]\} \ \cup\ \lft::\H_1 \ \cup\ \textsf{r}::\H_1\\
  \H_2 &=& \{[~]\} \ \cup\ \lft::\H_2 \ \cup\ \textsf{r}::\H_2\\
  \H_{\infty} &=& \{[~]\} \ \cup\ \lft::\H_\infty \ \cup\ \textsf{r}::\H_{\infty}
\end{eqnarray*}
Among all the fixpoints of the above equation, $\H_0$ is the least fixpoint and describes the inductive type associated with this equation, that is the type of the finite histories and
$\H_\infty$ is the greatest fixpoint and describes the coinductive type associated with this equation, that is the type of the infinite and infinite histories.  The principle that says
that given an equation, the least fixpoint is the inductive type associated with this equation and the greatest fixpoint is the coinductive type associated with this equation is very
general and will be used all along this paper. 

In the \Coq{} vernacular,  we are more talkative, but also more precise to describe the \textsf{CoInductive} type \emph{History}, (see appendix~\ref{sec:coq_vern} for the
definition.).  Recall the definition
\begin{quotation}
  A \textbf{coinductive} history (or a finite or infinite history) is
  \begin{itemize}
  \item either the empty history $[\ ]$,
  \item or a history of the form $c::h$, where $c$ is a choice and $h$ is  a history.
  \end{itemize}
\end{quotation}
which says that a \emph{history} is either the null history or a history which is a history, finite or infinite, increased by a choice at its head.  The word \textbf{coinductive} guarantees that we define actually infinite objects and attach to the objects of type
\emph{History} a specific form of reasoning, called \emph{coinduction}.  In coinduction, we assume that we ``know'' an infinite object by observing it through its definition, which
is done by a kind of peeling.  Since on infinite objects there is no concept of being smaller, one does not reason by saying ``I~know that the property holds on smaller objects let
us prove it on the object''.  On the contrary one says ``Let us prove a property on an infinite object, for that peel the object, assume that the property holds on the peeled
object and prove that it holds on the whole object''.  One does not say that the object is smaller, just that the property holds on the peeled object.  The above presentation is
completely informal, but it has been, formally founded in the theory of \Coq, after the pioneer works of David Park~\cite{DBLP:conf/tcs/Park81} and Robin Milner~\cite{Milner89},
using the concept of greatest fixpoint in type theory~\cite{DBLP:conf/types/Coquand93}.  Bertot and Cast\'eran~\cite{BertotCasterant04} present the concepts in Chapter~13 of their
book.

By just observing them, one cannot prove that two objects which have exactly the same behavior are equal, we can just say that they are observably equivalent.  Observable
equivalence is a relation weaker than equality\footnote{We are talking here about \emph{Leibniz equality}, not about \emph{extensional equality} see appendix~\ref{sec:eq}.}, called
\emph{bisimilarity} and defined on \emph{History} as a \textsf{CoInductive} (see appendix for a fully formal definition in the \Coq{ vernacular):

\begin{quotation}
  Bisimilarity $\hbis$ on \emph{histories} is defined \textbf{coinductively} as follows:
  \begin{itemize}
  \item $[\ ] \hbis [\ ]$,
  \item $h \hbis h'$ implies $\forall a:Agent, a :: h \hbis a::h'$.
  \end{itemize}
\end{quotation}

This means that two histories (why not infinite) are bisimilar if both are null or for composed histories, if both have the same head and the rests of both histories are bisimilar.
One can prove that two objects that are equal are bisimilar, but not the other way around, because for two objects to be equivalent by observation, does not mean that they have the
same structure. To illustrate the difference between bisimilarity and equality of infinite objects let us consider for example two infinite histories $`a_0$ and $`b_0$ that are
obtained as solutions of two equations, more precisely as two cofixpoints.  Let $`a_n = c(n):: `a_{n+1}$, where $c(n)$ is
$(\textbf{if}~even(n)~\textbf{then}~\lft~\textbf{else}~\textsf{r})$, and $`b_p$ is $\lft :: \textsf{r} :: `b_{p+1}$.  We know that if we ask for the $5^{th}$ element of $`a_0$ and
$`b_0$ we will get $\lft$ in both cases, and the $2p^{th}$ element will be $\textsf{r}$ in both cases, but we have no way to prove that $`a_0$ and $`b_0$ are equal, i.e., have
exactly the same structure.  Actually the picture in Figure~\ref{fig:a0b0} shows that they look different and there is no hope to prove by induction, for instance, that they are
the same, since they are not well-founded.

\begin{figure}[t]
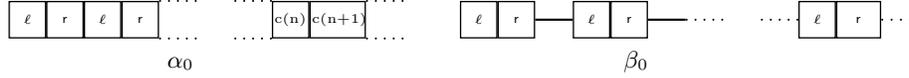

  \bigskip \bigskip

\noindent
 \psframe[linewidth=.5pt,fillstyle=solid](0,0)(.5,.5)
\psframe[linewidth=.5pt,fillstyle=solid](.5,0)(1,.5) %
\psframe[linewidth=.5pt,fillstyle=solid](1,0)(1.5,.5) %
\psframe[linewidth=.5pt,fillstyle=solid](1.5,0)(2,.5) %
\rput(.25,.25){\tiny \lft} %
\rput(.75,.25){\tiny \textsf{r}} %
\rput(1.25,.25){\tiny \lft} %
\rput(1.75,.25){\tiny \textsf{r}} %
\psline[linestyle=dotted](2,.5)(2.5,.5) %
\psline[linestyle=dotted](2,0)(2.5,0) %
\psline[linestyle=dotted](3,.5)(3.5,.5)
\psline[linestyle=dotted](3,0)(3.5,0) %
\psframe[linewidth=.5pt,fillstyle=solid](3.5,0)(4,.5)
\psframe[linewidth=.5pt,fillstyle=solid](4,0)(4.75,.5) %
\rput(3.75,.25){\tiny c(n)}
\rput(4.4,.25){\tiny c(n+1)}
\psline[linestyle=dotted](4.75,.5)(5.25,.5)
\psline[linestyle=dotted](4.75,0)(5.25,0) %
\psframe[linewidth=.5pt,fillstyle=solid](6,0)(6.5,.5)%
\psframe[linewidth=.5pt,fillstyle=solid](6.5,0)(7,.5) 
\rput(6.25,.25){\tiny \lft}
\rput(6.75,.25){\tiny \textsf{r}}
\psline(7,.25)(7.5,.25) %
\psframe[linewidth=.5pt,fillstyle=solid](7.5,0)(8,.5)%
\psframe[linewidth=.5pt,fillstyle=solid](8,0)(8.5,.5) 
\rput(7.75,.25){\tiny
\lft}
\rput(8.25,.25){\tiny \textsf{r}} 
\psline(8.5,.25)(9,.25) 
\psline[linestyle=dotted](9,.25)(9.5,.25) 
\psline[linestyle=dotted](10,.25)(10.5,.25)
\psframe[linewidth=.5pt,fillstyle=solid](10.5,0)(11,.5)%
\psframe[linewidth=.5pt,fillstyle=solid](11,0)(11.6,.5) 
\rput(10.75,.25){\tiny \lft}
\rput(11.3,.25){\tiny \textsf{r}}
\psline[linestyle=dotted](11.6,.25)(12,.25)

\hspace*{2cm} $`a_0$ \hspace*{5.5cm} $`b_0$
  \caption{The picture of two bisimilar histories}
  \label{fig:a0b0}
\end{figure}

\section{Infinite Games}
\label{sec:games}

In classical textbooks, finite and infinite games are presented through their histories. But in the framework of a proof assistant or just to make rigorous proofs, it makes sense
to present them structurally.  Therefore, games are rather naturally seen as either a leaf to which a \emph{utility function} (a function that assigns a utility to each agent, aka
an \emph{outcome}) is attached or a node which is associated to an agent and two subgames.  If agents are \emph{Alice} and \emph{Bob} and utilities are natural numbers, a utility
function can be the function $\{\textit{Alice}"|->" 3,  \textit{Bob} "|->" 2\}$.
In Section~\ref{sec:myria}, we focus on \emph{centipede games} which are games in which the
right part is always a leaf and in which the left subgame is always a composed game.  Actually we study games that ``can'' be infinite and ``can'' have finite or infinite branches.

\begin{quotation}
  The type of \emph{Game}s is defined as a \textbf{coinductive} as follows:
  \begin{itemize}
  \item a \emph{Utility function} makes a \emph{Game},
  \item an \emph{Agent} and two \emph{Game}s make a \emph{Game}.
  \end{itemize}
\end{quotation}

A \emph{Game} is either a leaf (a terminal node) or a composed game made of an agent (the agent who has the turn) and two subgames (the formal definition in the \Coq{} vernacular
is given in the appendix~\ref{sec:coq_vern}).  We use the expression \coqdocid{gLeaf} $f$ to denote the leaf game associated with the utility function $f$ and the expression
\coqdocid{gNode} $a$ $g_l$ $g_r$ to denote the game with agent $a$ at the root and two subgames $g_l$ and $g_r$.
For instance, the game we would draw:
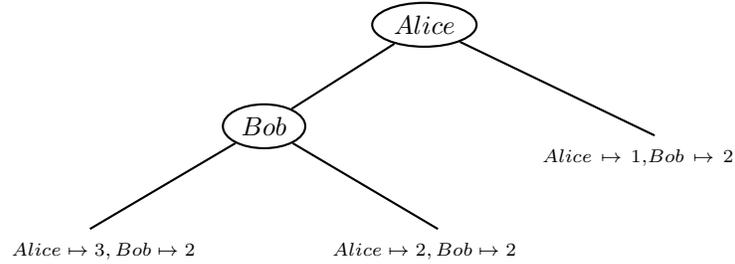
\label{pag:game}

\[\begin{psmatrix}[colsep=10pt,rowsep=20pt]
  &&&{\ovalnode{a}{Alice}} \\
  &&{\ovalnode{b}{Bob}} && [name=c]{\raisebox{-10pt}{$\scriptstyle Alice ~\mapsto~ 1, Bob ~\mapsto~ 2$}}\\
  &[name=d]{\raisebox{-10pt}{\scriptsize $Alice \mapsto 3, Bob \mapsto 2$}} && [name=e]{\raisebox{-10pt}{\scriptsize$Alice \mapsto 2, Bob \mapsto 2$}}
  \ncline{a}{b}
  \ncline{a}{c}
  \ncline{b}{d}
  \ncline{b}{e}
\end{psmatrix}
\]

\medskip

is represented by the term:

  (\coqdocid{gNode} \emph{Alice} (\coqdocid{gLeaf} $\scriptstyle Alice ~\mapsto~ 1, Bob ~\mapsto~ 2$ )\\
  \hspace*{70pt} 
  (\coqdocid{gNode} \emph{Bob} 
  (\coqdocid{gLeaf} {\scriptsize $Alice \mapsto 2, Bob \mapsto 2$}) 
  (\coqdocid{gLeaf} {\scriptsize$Alice \mapsto 3, Bob \mapsto 2$}))\\

Hence one builds a finite game in two ways: either
a given utility function $f$ is encapsulated by the operator \coqdocid{gLeaf} to make the game (\coqdocid{gLeaf} \coqdocid{f}), or an agent \coqdocid{a} and two games $g_l$ and
$g_r$are given to make the game (\coqdocid{gNode} \coqdocid{a} $g_l$ $g_r$).  Notice that in such games, it can be the case that the same
agent \coqdocid{a} has the turn twice in a row, like in the game (\coqdocid{gNode} \coqdocid{a} (\coqdocid{gNode} \coqdocid{a} $g_1$ $g_2$) $g_3$}).

Concerning comparisons of utilities we consider a very general setting where a utility is no more that a type (a ``set'') with a preference which is a
preorder, i.~e., a transitive and reflexive relation, and which we write \leut. A preorder is enough for what we want to prove.
We assign to the leaves, a utility function which associates a utility to each agent.

We can also tell how we associate a history with a game or a history and a utility function with a game (see the \Coq{} script).  We will see in the next section how to associate a
utility with an agent in a game, this is done in the frame of a strategy, which is described now.

\section{Infinite Strategies}
\label{sec:inf_stra}

The main concept of this paper is this of infinite strategy\footnote{Authors call it a \emph{strategy profile}, but in this paper, following Vestergaard~\cite{vestergaard06:IPL},
  we call it a \emph{strategy} for convenience in the formal development and we give it a formal definition.} which is a coinductive.  More specifically, in this paper, we focus on
infinite binary strategies associated with infinite binary games.
\begin{quotation}
  The type of \emph{Strategies} is defined as a \textbf{coinductive} as follows:
  \begin{itemize}
  \item a \emph{Utility function} makes a \emph{Strategy}.
  \item an \emph{Agent}, a \emph{Choice} and two \emph{Strategies} make a \emph{Strategy}.
  \end{itemize}
\end{quotation}

Basically\footnote{The formal definition in the \Coq{} vernacular is given in appendix~\ref{sec:coq_vern}.} an infinite strategy which is not a leaf is a node with four
items: an agent, a choice, two infinite strategies.  A strategy is the same as a game, except that there is a choice.  In what follows, since we consider equilibria, we only
address strategies.  Strategies of the first kind are written $\og f \fg$ and strategies of the second kind are written $\og a,c,s_l,s_r\fg$.  In other words, if between the ``$\og$''
and the ``$\fg$'' there is one component, this component is a utility function and the result is a leaf strategy and if there are four components, this is a node strategy.  For
instance, with the game of page~\pageref{pag:game} one can associate at least the following strategies:

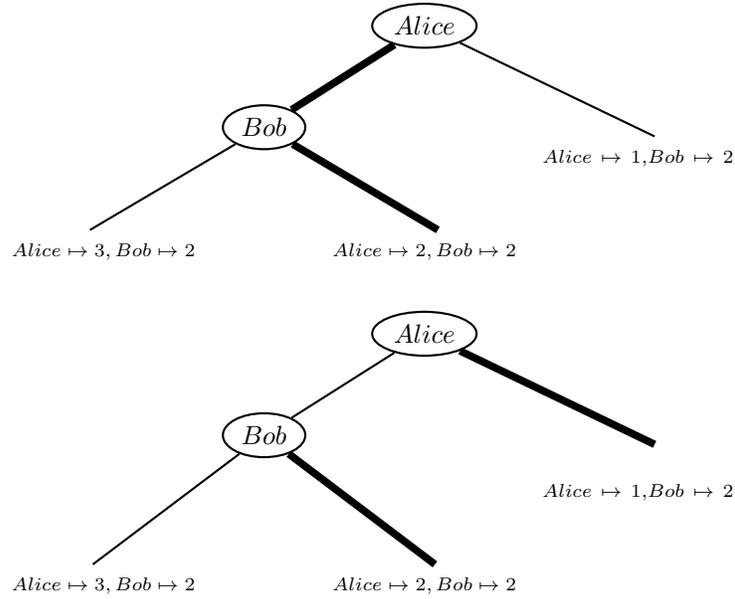
\[\begin{psmatrix}[colsep=10pt,rowsep=20pt]
  &&&{\ovalnode{a}{Alice}} \\
  &&{\ovalnode{b}{Bob}} && [name=c]{\raisebox{-10pt}{$\scriptstyle Alice ~\mapsto~ 1, Bob ~\mapsto~ 2$}}\\
  &[name=d]{\raisebox{-10pt}{\scriptsize $Alice \mapsto 3, Bob \mapsto 2$}} && [name=e]{\raisebox{-10pt}{\scriptsize$Alice \mapsto 2, Bob \mapsto 2$}}
  \ncline[linewidth=.1]{a}{b}
  \ncline{a}{c}
  \ncline{b}{d}
  \ncline[linewidth=.1]{b}{e}
\end{psmatrix} %
\]
\[
\begin{psmatrix}[colsep=10pt,rowsep=20pt]
  &&&{\ovalnode{a}{Alice}} \\
  &&{\ovalnode{b}{Bob}} && [name=c]{\raisebox{-20pt}{$\scriptstyle Alice ~\mapsto~ 1, Bob ~\mapsto~ 2$}}\\
  &[name=d]{\raisebox{-10pt}{\scriptsize $Alice \mapsto 3, Bob \mapsto 2$}} && [name=e]{\raisebox{-10pt}{\scriptsize$Alice \mapsto 2, Bob \mapsto 2$}}
  \ncline{a}{b}
  \ncline[linewidth=.1]{a}{c}
  \ncline{b}{d}
  \ncline[linewidth=.1]{b}{e}
\end{psmatrix}
\]

\medskip
\noindent which correspond to the expressions

  $\og$\emph{Alice},\textsf{r},$\og$ $\scriptstyle Alice ~\mapsto~ 1, Bob ~\mapsto~ 2$ $\fg$,\\
  \hspace*{65pt} $\og$\emph{Bob},\lft,$\og${\scriptsize $Alice \mapsto 2, Bob \mapsto 2$}$\fg$,$\og${\scriptsize$Alice \mapsto 3, Bob \mapsto
    2$}$\fg$ $\fg$ 

and

  $\og$\emph{Alice},\lft,$\og$ $\scriptstyle Alice ~\mapsto~ 1, Bob ~\mapsto~ 2$ $\fg$,\\
  \hspace*{65pt} $\og$\emph{Bob},\lft,$\og${\scriptsize $Alice \mapsto 3, Bob \mapsto 2$}$\fg$,$\og${\scriptsize$Alice \mapsto 2, Bob \mapsto
    2$}$\fg$,
  $\fg$.

To describe an infinite strategy one uses most of the time a fixpoint equation like:
\[t \quad = \quad \og Alice, \lft, \og {\scriptstyle Alice ~\mapsto~ 0, Bob ~\mapsto~ 0}\fg, \og Bob, \lft, t, t\fg\fg\]
which correspond to the pictures:
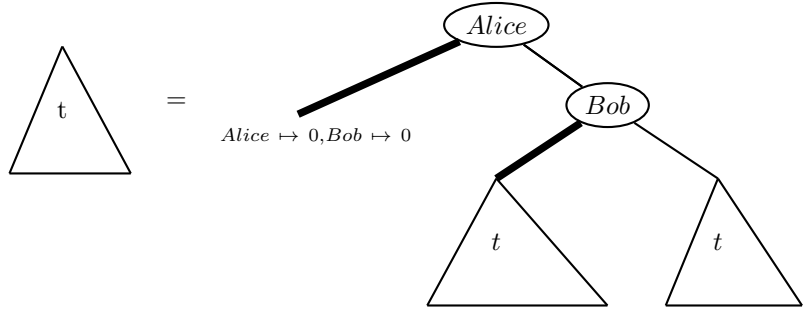
\[
\raisebox{50pt}{
  \begin{psmatrix}[colsep=6pt,rowsep=12pt]
  &&&&[name=0]\\
  &&&&t\\
  &[name=1]&&&&&&&[name=2]
  \ncline{0}{1}
  \ncline{0}{2}
  \ncline{1}{2}
\end{psmatrix}
}
\quad \raisebox{75pt}{=} \ \
\begin{psmatrix}[colsep=6pt,rowsep=12pt]
  &&& {\ovalnode{sommet}{Alice}}\\
  & [name=1]{\raisebox{-10pt}{$\scriptstyle Alice ~\mapsto~ 0, Bob ~\mapsto~ 0$}} &&& {\ovalnode{2}{Bob}}\\
  &&&[name=21]&&&&& [name=22]\\
  &&& t &&&&& t\\
  && [name=211] && [name=212] & [name=221] &&&&&&&& [name=222] & 
  \ncline[linewidth=.1]{sommet}{1}
  \ncline{sommet}{2}
  \ncline[linewidth=.1]{2}{21}
  \ncline{sommet}{2}
  \ncline{2}{22}
  \ncline{21}{211}
  \ncline{21}{212}
  \ncline{211}{212}
  \ncline{22}{221}
  \ncline{22}{222}
  \ncline{221}{222}
\end{psmatrix}
\]

Other examples of infinite strategies are given in Section~\ref{sec:inf_ped}.  Usually an infinite game is defined as a cofixpoint, \ie as the solution of an equation, possibly a parametric equation.

Whereas in the finite case one can easily associate with a strategy a utility function, \ie a function which assigns a utility to an agent, as the result of a recursive evaluation,
this is no more the case with infinite strategies.  One reason is that one is not sure that such a utility function exists for the strategy.  This makes the function partial, which
cannot be defined as inductive or coinductive.  Therefore $s2u$ (an abbreviation for \emph{strategy-to-utility}) is a relation between a strategy and
a utility function, which is also a coinductive; $s2u$ appears in expression of the form $(s2u\ s\ a\ u)$ where $s$ is a strategy, $a$ is an agent and $u$ is a utility.
It reads ``$u$ is a utility of the agent $a$ in the strategy~$s$''.

\begin{quotation}
$s2u$ is a predicate defined \textbf{coinductively} as follows:
  \begin{itemize}
  \item $s2u \og f \fg~a~(f(a))$ holds,
  \item if $s2u~s_l~a~u$ holds then $s2u~\og a',\lft,s_l,s_r\fg~a~u$ holds,
  \item if $s2u~s_r~a~u$ holds then $s2u~\og a',\textsf{r},s_l,s_r\fg~a~u$ holds.
  \end{itemize} 
\end{quotation}

This means the utility of $a$ for the leaf strategy $\og f\fg$ is $f(a)$, \ie the value delivered by the function $f$ when applied to $a$.  The utility of $a$ for the strategy
$\og a',\lft,s_l,s_r\fg$ is $u$ if the utility of $a$ for the strategy $s_l$ is $u$. If one call $s_0$ the first above strategy, one
has $s2u~s_0~\emph{Alice}~2$, which means that, for the strategy~$s_0$, the utility of \emph{Alice} is $2$.
In order to insure that $s2u$ has a result we define an operator
\ltl{} that says that if one follows the choices shown by the strategy one reaches a leaf, i.e., one does not go forever.

\begin{quotation}
The predicate \ltl{} is defined \textbf{inductively} as
  \begin{itemize}
  \item the strategy $\og f \fg$ \ltl{},
  \item if $s_l$ \ltl{}, then $\og a, \mathsf{l}, s_l, s_r\fg$ \ltl{},
  \item if $s_r$ \ltl{}, then $\og a, \mathsf{r}, s_l, s_r\fg$ \ltl{}.
  \end{itemize}
\end{quotation}

This means that a strategy which is itself a leaf \ltl{} and if the strategy is a node, if the choice is \lft{} and if the left substrategy 
\ltl{} then the whole strategy \ltl{} and similarly if the choice is \textsf{r}.  We claim that this gives a good notion of \emph{finite horizon} which seems to
be rather a concept on strategies than on games.

If $s$ is a strategy that satisfies the predicate \ltl{} then the utility exists and is unique, in other words:
\begin{itemize}
\item[$`(!)$] For all agent $a$ and for all strategy $s$, if $s$ \ltl{} then there exists a utility $u$ which ``is a utility of the agent $a$ in the strategy~$s$''.
\item[$`(!)$] For all agent $a$ and for all strategy $s$, if $s$ \ltl{}, if ``$u$ is a utility of the agent $a$ in the strategy~$s$'' and ``$v$ is a utility of the agent
  $a$ in the strategy~$s$'' then $u=v$.
\end{itemize}

We also consider a predicate \altl{} which means that everywhere in the strategy, if one follows the choices, one leads to a leaf.  This property is defined everywhere on an
infinite strategy and is therefore coinductive.

\begin{quotation}
  The predicate \altl{} is defined \textbf{coinductively}
  \begin{itemize}
  \item the strategy  $\og f \fg$ \altl{},
  \item for all choice $c$, if $\og a, \mathsf{r}, s_l, s_r\fg$ \ltl{},  if $s_l$ \altl{}, if $s_r$ \altl{}, then $\og a,
    \mathsf{r}, s_l, s_r\fg$ \altl{}. 
  \end{itemize}
\end{quotation}

This says that a strategy, which is a leaf, \altl{} and that a composed strategy inherits the predicate from its substrategies provided itself \ltl{}.  We
define also bisimilarity between games and between strategies.  For strategies, this is defined by:\pagebreak[3]

\begin{quotation}
  The bisimilarity $\sbis$ on \emph{strategies} is defined \textbf{coinductively} as follows:
  \begin{itemize}
  \item $\og f\fg \ \sbis\ \og f\fg$,
  \item if $s_l\sbis s_l'$ and $s_r \sbis s_r'$ then $\og a, c, s,_l s_r\fg\ \sbis\ \og a,c, s_l', s_r'\fg$.
  \end{itemize}
\end{quotation}

This says that two leaves are bisimilar if and only if they have the same utility function and that two strategies are bisimilar if and only if they have the same head agent, the same choice
and bisimilar substrategies.

\section{Histories of strategies and finite strategies}
\label{sec:bis}

In this short section, we present concepts that are not directly connected with equilibria, but that are of interest.  For a finite or infinite history to be one of the histories
of a game is characterized by a predicate.  We define also the history of a strategy which is the history that consists in following the choices and which is typically obtained by
a function.  This function is a function on infinite objects since it associates an infinite object (a history) with an infinite object (a strategy).  This kind of functions are
defined as cofixpoints and are specific to infinite objects.  Without surprise, the history of a strategy ``is a history of'' the game associated with that strategy.  For a
strategy to be finite is characterized by a unary predicate.  If the history of a strategy is infinite then the strategy itself is infinite.

\section{Convertibility}
\label{sec:conv}

To speak about equilibria, one needs to speak first about a relation between strategies which is called \emph{convertibility}.  In classical game theory, a strategy $s$ is given
and a convertible strategy $(r_a,s_{-a})$ is the strategy in which agent $a$ chooses $r_a$ while other agents $b$ choose $s_b$.  We call the relation that relates $s$ with
$(r_a,s_{-a})$ \emph{convertibility} and we write it $"|-|"$.  Basically two convertible strategies have the same ``skeleton'', \ie the same underlying structure.  

\begin{quotation}
  The relation $\conva$ is defined \textbf{inductively} as follows:
  \begin{itemize}
  \item $\conva$ is reflexive, i.e., for all $s$, $s \conva s$.
  \item If the node has the same agent as the agent in $\conva$ then the choice may change, i.e.,
    \[
    \prooftree s_1 \conva s_1'\qquad s_2 \conva s_2' %
    \justifies \og a, c, s_1, s_2 \fg \ \conva \ \og a, c', s_1', s_2'\fg
    \endprooftree
    \]

  \item If the node does not have the same agent as in $\conva$, then the choice has to be the same:
    \[
    \prooftree 
  s_1 \conva s_1'\qquad s_2 \conva s_2' %
  \justifies \og a', c, s_1, s_2 \fg \ \conva \ \og a', c, s_1', s_2'\fg
  \endprooftree
  \]
\end{itemize}
\end{quotation}
Since the relation is defined inductively it compares only a finite part of the strategies.  Moreover convertibility preserves the predicate \coqdocid{AlwLeadsToLeaf}.
This is illustrated by the following lemma:

\begin{quotation}
  For all $a$, $s$, $s'$, if $s$ \altl{} and $s \conva s''$ then $s'$ \altl.
\end{quotation}

Actually there is another convertibility which we write $\coconva$:
\begin{quotation}
The relation $\coconva$ is defined \textbf{coinductively} as follows:
  \begin{itemize}
  \item A leaf is only convertible to itself.
  \item For nodes, one has basically the rules:
    \[
    \prooftree s_1 \coconva s_1'\qquad s_2 \coconva s_2' %
    \justifies \og a, c, s_1, s_2 \fg \ \coconva \ \og a, c', s_1', s_2'\fg
    \endprooftree
    \]
    \[
    \prooftree 
  s_1 \coconva s_1'\qquad s_2 \coconva s_2' %
  \justifies \og a', c, s_1, s_2 \fg \ \coconva \ \og a', c, s_1', s_2'\fg
  \endprooftree
  \]
\end{itemize}
\end{quotation}

From their definition, one can prove that both convertibilities are equivalence relations.  The difference between the two convertibilities appears only in the \Coq{} script and
shows typically the difference between induction and coinduction.  It can be sketched as follows: $\coconva$ traverses the whole strategies and can accept infinitely many changes
in choices, whereas $\conva$ inspects only finite parts of the strategies, \ie finitely many choices.  We claim that human agents apply the convertibility $\conva$ as they can only
carry a finite amount of reasoning and comparison.  Notice that when two strategies are convertible there underlying games are bisimilar.

\section{Equilibria}
\label{sec:equilib}

\subsection{Nash equilibria}
\label{sec:NE}

The notion of Nash equilibrium is translated from the notion in textbooks.   Since we have two concepts of convertibility, we might have two concepts of Nash equilibria,
but as said we focus only on the inductive notion of convertibility $\conva$. The concept of Nash equilibrium is based on a
comparison of utilities; this assumes that an actual utility exists and therefore this requires convertible strategies to \textit{``lead to a leaf''}.
$s$ is a \emph{Nash equilibrium} if the following implication holds: if $s$ \ltl{} and for all agent~$a$ and for all strategy~$s'$ which is convertible to $s$,
i.e., $s\conva s'$, and which \ltl{}, if $u$ is the utility of~$s$ for~$a$ and $u'$ is the utility of $s$' for $a$, then $u' \leut u$.  




\subsection{Subgame Perfect Equilibria}
\label{sec:sgpe}

Let us consider now \emph{subgame perfect equilibria}, which we write $SGPE$.  $SGPE$ is a property of strategies. It requires the substrategies to fulfill coinductively the same
property, namely to be a $SGPE$, and to insure that the strategy with the best utility for the node agent to be chosen.  Since both the strategy and its substrategies are infinite,
it makes sense to make the definition of $SGPE$ coinductive.   

\begin{quotation}
  $SGPE$ is defined \textbf{coinductively} as follows:
  \begin{itemize}
  \item $SGPE \og f\fg$,
  \item if $\og a,\lft,s_l,s_r\fg$ \altl{}, if $SGPE(s_l)$ and $SGPE(s_r)$, if $s2u~s_l~a~u$ and $s2u~s_r~a~v$, if $v\le u$ then $SGPE~\og a,\lft,s_l,s_r\fg$,
  \item if $\og a,\textsf{r},s_l,s_r\fg$ \altl{}, if $SGPE(s_l)$ and $SGPE(s_r)$, if $s2u~s_l~a~u$ and $s2u~s_r~a~v$, if $u\le v$ then $SGPE~\og a,\textsf{r},s_l,s_r\fg$,
  \end{itemize}
\end{quotation}

It means that a strategy, which is a
leaf, is a subgame perfect equilibrium (condition \emph{SGPE\_Leaf}).  Moreover if the strategy is a node, if the strategy \altl{}, if it has agent $a$ and
choice \lft, if both substrategies are subgame perfect equilibria and if the utility of the agent $a$ for the right substrategy is less than this for the
left substrategy then the whole strategy is a subgame perfect equilibrium and vice versa (condition \emph{SGPE\_left}).  If the choice is \textsf{r} (condition
\emph{SGPE\_right}) this works similarly.

Notice that since we require that the utility can be computed not only for the strategy, but for the substrategies and for the subsubstrategies and so on, we require these strategies no
only to \textit{``lead to a leaf''} but to \textit{``always lead to a leaf''}.

\section{Two infinite games with centipede shape}
\label{sec:inf_ped}

In this section we study two kinds of games that have some analogies, especially they have a centipede shape, since they have an infinite backbone (on the ``left'') and all the
right subgames are leaves.  In both cases, the utilities go to infinity, but in the first (dollar auction game) the utilities go to $(-\infty,-\infty)$ (costs, \ie the opposites of
utilities, go to $(+\infty,+\infty)$), whereas in the second, (centipede game) the utilities go to $(+\infty,+\infty)$.  One common mistake in analyzing those games is to believe
that results on the infinite game can be obtain as the limit of partial results on the finite games, which is not the case (see~\cite{fagin93:_finit_model_theor_person_persp} for
the perspective of a model theorist).

\subsection{The ``Illogic of Conflict Escalation''}
\label{sec:escal}

\begin{figure}[t]
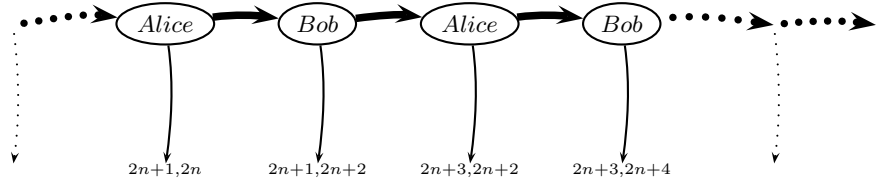

  \centering
\hspace*{-30pt} 
\(\begin{psmatrix}[colsep=20pt]
  & [name=o]
  &{\ovalnode{a}{Alice}} &{\ovalnode{b}{Bob}} & {\ovalnode{a1}{Alice}} &{\ovalnode{b1}{Bob}}  & [name=c] & [name=d]
  &\\
  &[name=p]\phantom{\scriptstyle 2n+1, 2n+2} 
  &[name=e] {\scriptstyle 2n+1, 2n} 
  &[name=f] {\scriptstyle 2n+1, 2n+2} %
  &[name=e1] {\scriptstyle 2n+3, 2n+2} 
  &[name=f1] {\scriptstyle 2n+3, 2n+4} %
  &[name=h] \phantom{\scriptstyle 2n+1, 2n+2} 
  \ncarc[arrows=->,linestyle=dotted]{o}{p} %
  \ncarc[arrows=->,linestyle=dotted,linewidth=.1]{o}{a} %
  \ncarc[arrows=->,linewidth=.1]{a}{b} %
  \ncarc[arrows=->,linewidth=.1]{b}{a1} %
  \ncarc[arrows=->,linewidth=.1]{a1}{b1} %
  \ncarc[arrows=->,linestyle=dotted,linewidth=.1]{b1}{c} %
  \ncarc[arrows=->,linestyle=dotted,linewidth=.1]{c}{d} %
  \ncarc[arrows=->]{a}{e} %
  \ncarc[arrows=->]{b}{f} %
  \ncarc[arrows=->]{a1}{e1} %
  \ncarc[arrows=->]{b1}{f1} %
  \ncarc[arrows=->,linestyle=dotted]{c}{h} %
\end{psmatrix}
\)
  \caption{The strategy  ``never give up'' for the \emph{dollar auction game} (payoffs are $100v$ minus the numbers)}
  \label{fig:agu}
\end{figure}

\begin{figure}[b]
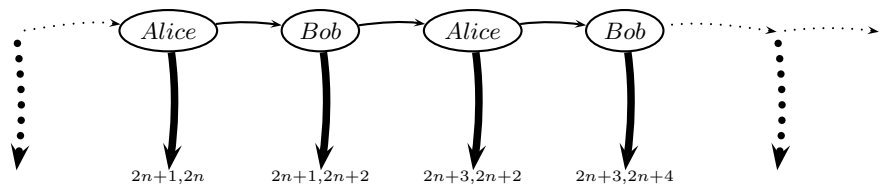

  \centering
  \bigskip
  \hspace*{-30pt} 
\(\begin{psmatrix}[colsep=20pt]
  & [name=o]
  &{\ovalnode{a}{Alice}} &{\ovalnode{b}{Bob}} & {\ovalnode{a1}{Alice}} &{\ovalnode{b1}{Bob}}  & [name=c] & [name=d]
  &\\
  &[name=p]\phantom{\scriptstyle 2n+1, 2n+2} 
  &[name=e] {\scriptstyle 2n+1, 2n} 
  &[name=f] {\scriptstyle 2n+1, 2n+2} %
  &[name=e1] {\scriptstyle 2n+3, 2n+2} 
  &[name=f1] {\scriptstyle 2n+3, 2n+4} %
  &[name=h] \phantom{\scriptstyle 2n+1, 2n+2} 
  \ncarc[arrows=->,linestyle=dotted,linewidth=.1]{o}{p} %
  \ncarc[arrows=->,linestyle=dotted]{o}{a} %
  \ncarc[arrows=->]{a}{b} %
  \ncarc[arrows=->]{b}{a1} %
  \ncarc[arrows=->]{a1}{b1} %
  \ncarc[arrows=->,linestyle=dotted]{b1}{c} %
  \ncarc[arrows=->,linestyle=dotted]{c}{d} %
  \ncarc[arrows=->,linewidth=.1]{a}{e} %
  \ncarc[arrows=->,linewidth=.1]{b}{f} %
  \ncarc[arrows=->,linewidth=.1]{a1}{e1} %
  \ncarc[arrows=->,linewidth=.1]{b1}{f1} %
  \ncarc[arrows=->,linewidth=.1,linestyle=dotted]{c}{h} %
\end{psmatrix}
\)

\bigskip
  \caption{The strategy ``always give up'' for the \emph{dollar auction game} (payoffs are $100v$ minus the numbers)}
  \label{fig:ngu}
\end{figure}

The above development has been applied to an example called the \emph{dollar auction game} by Shubik~\cite{Shubik:1971} and the \emph{illogic conflict of escalation} by
Gintis~\cite{gintis00:_game_theor_evolv}.  Recall its principle. Two agents \emph{Alice} and \emph{Bob} compete in an auction for an object of a value $v\,\$$
($100v$\,\textcent).  The two agents bid $2$\,\textcent, one after the other.  We assume that there is no limit on the bids, \ie the bankroll of the agents is not limited.  This
generalization is possible because we reason on infinite games.  By the way, if there would be a limit on the bids, there would be no escalation.  If one agent gives up, the highest
bidder gets the object, but the second bidder pays also for his (her) bid.  As Shubik noted, this game may never cease, due to what
Colman~\cite{colman99:_game_theor_and_its_applic} calls the\emph{ Macbeth effect}.  We have closely studied only two infinite strategies: a strategy called \emph{always give up}
($agu$ in short) (see Figure~\ref{fig:agu}) and a strategy called \emph{never give up} ($ngu$ in short) (see Figure~\ref{fig:ngu}).  In the strategy \emph{agu}, both agents leave
the auction at each step, when in the strategy \emph{ngu}, the agents keep bidding, this is the escalation.  \emph{agu} and \emph{ngu} are solutions of parametric equations.  For
instance \emph{agu} is solution of the equation (cofixpoint):
\[ agu_n \ = \ \og Alice, \lft ,\og Bob, \lft,agu_{n+1},[2n+1, 2n+2]\fg,[2n+1,2n]\fg.\]
where $[x,y]$ stands for $\og Alice "|->" x, Bob "|->" y\fg$.
We are able to prove in \Coq{} that $agu$ is a \emph{subgame perfect
  equilibria} and that both $agu$ and $ngu$ are Nash equilibria.  The proof of $agu$ being a subgame perfect equilibrium is based on a coinductive argument, whereas the proof of
$agu$ being a Nash equilibrium is by case; $ngu$ is a Nash equilibrium because it does not ``lead to a leaf'', this last fact comes from a coinductive argument.

\subsection{Centipede}
\label{sec:myria}

\begin{figure}[t]
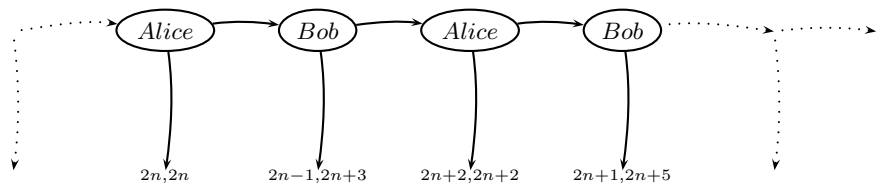

  \centering
  \bigskip
  \hspace*{-30pt} 
\(\begin{psmatrix}[colsep=20pt]
  & [name=o]
  &{\ovalnode{a}{Alice}} &{\ovalnode{b}{Bob}} & {\ovalnode{a1}{Alice}} &{\ovalnode{b1}{Bob}}  & [name=c] & [name=d]
  &\\
  &[name=p]\phantom{\scriptstyle 2n+1, 2n+2} 
  &[name=e] {\scriptstyle 2n, 2n} 
  &[name=f] {\scriptstyle 2n-1, 2n+3} %
  &[name=e1] {\scriptstyle 2n+2, 2n+2} 
  &[name=f1] {\scriptstyle 2n+1, 2n+5} %
  &[name=h] \phantom{\scriptstyle 2n+1, 2n+2} 
  \ncarc[arrows=->,linestyle=dotted]{o}{p} %
  \ncarc[arrows=->,linestyle=dotted]{o}{a} %
  \ncarc[arrows=->]{a}{b} %
  \ncarc[arrows=->]{b}{a1} %
  \ncarc[arrows=->]{a1}{b1} %
  \ncarc[arrows=->,linestyle=dotted]{b1}{c} %
  \ncarc[arrows=->,linestyle=dotted]{c}{d} %
  \ncarc[arrows=->]{a}{e} %
  \ncarc[arrows=->]{b}{f} %
  \ncarc[arrows=->]{a1}{e1} %
  \ncarc[arrows=->]{b1}{f1} %
  \ncarc[arrows=->,linestyle=dotted]{c}{h} %
\end{psmatrix}
\)
\bigskip
  \caption{The centipede game}
  \label{fig:cent}
\end{figure}

The centipede games have been introduced by~\cite{rosenthal81:_games_of_perfec_infor_predat} (see also~\cite{Colman_rat_back_ind,osborne94:_cours_game_theory}), but only as finite
games.  The paradox is that the subgame perfect equilibrium is when the player do not enter the game, whereas clearly if they would enter the game, they would get a better payoff.
In this development, we consider an infinite game and show that the strategy that never stops is also a Nash equilibrium; in other words if players enter the game they should not stop.  Payoffs
are given in Figure~\ref{fig:cent}.  Like for the \emph{dollar auction game}, we consider two strategies, namely a strategy \emph{always give up} and a strategy \emph{never give
  up} and we can prove by coinduction that the first is both a subgame perfect equilibrium and a Nash equilibrium and that the second is a Nash equilibrium.  The proofs are very
similar to this for the \emph{dollar auction game}.  The strategy \emph{never give up} is a Nash equilibrium, for the same reason: the agent cannot compute the payoff, but there is a
difference, whereas the payoff cannot be computed in the \emph{dollar auction game} because it goes to $-\infty$, in the case of the \emph{centipede game} it cannot be computed
because it goes to $+\infty$. 

\subsection{Human and formal reasoning}
\label{hum_form}

Literature (see~\cite{colman99:_game_theor_and_its_applic} Section~9.3 for a detailed review) cites many experiments which have been used to compare human reasoning and formal
reasoning and in the so called \emph{illogic conflict of escalation} like in the \emph{centipede game}, it is advocated that humans are illogic (see for
instance~\cite{oneill86:_inten_escal_and_dollar_auction,leininger89:_escal_and_cooop_in_confl_situat}). Our work shows that this is not the case as we are able to prove that the
infinite escalation is also a Nash equilibrium.\footnote{We have not been able to prove formally in \Coq{} that \emph{always give up} and \emph{never give up} are the only Nash equilibria, but
  this seems very likely.}  In other words when an agent decides to enter the game, then he (she) has no other rational solution but to bid forever (\emph{Macbeth effect}) and
acting so makes perfect sense from the theoretical point of view of coinduction\footnote{We propose to call this the \emph{Jourdain effect}, from the character of
  \emph{Moli\`ere}'s play the \emph{Bourgeois genthilhomme} (The middle-class gentleman), delighted to learn that he has been speaking prose all his life without knowing it.
  \emph{``For more than forty years I have been speaking prose without knowing anything about it.''}(act II, scene 5)~\cite{wiki:MCG}.}, despite when it applies to war, it is
scaring.  So the advice should be: ``Never enter such a game, otherwise logic will push you to escalation''.
In games involving money, he (she) will then face bankruptcy, but this is another story.  In both games, assuming that the agents commonly believe that the game is
infinite, they have two possibilities, either not entering the game or entering it and running forever as both strategies are Nash equilibria.

\section{Coinduction in \Coq}
\label{sec:coq}

As we have said the proof assistant \Coq~\cite{Coq:manual} plaid a central role in this research. 

\subsubsection{Why should we formalize a concept in a proof assistant?}

To answer this question we like to cite Donald Knuth~\cite{shustek08:_inter_donal_knuth}:
\begin{it}
  \begin{quotation}
    People have said you don't understand something until you've taught it in a class.
    The truth is you don't really understand something until you've taught it to a computer, until you've been able to program it.
  \end{quotation}
\end{it}
We claim that we can appropriately replace the last sentence by \emph{``until you've taught it to a proof assistant, until you've code it into \Coq~\cite{Coq:manual},
  Isabelle~\cite{Nipkow-Paulson-Wenzel:2002}, or PVS~\cite{cade92-pvs}''} as it seems even more demanding to ``teach'' a proof assistant like \Coq{} than to write a program on the
same topics.  Actually without \Coq{}, we would not have been able to capture the concepts of Nash equilibrium and Subgame Perfect equilibria presented in this paper. This is
indeed the result of formal deduction, intuition and try and error in \Coq{} since proving properties of infinite games and infinite strategy profiles is extremely subtle (see
Section~\ref{sec:coq}).  Moreover by relying on a proof assistant, we can free this article from formal developments and tedious and detailed proofs, knowing anyway that they are
correct in any detail and that the reader will refer to the \Coq{} script in case of doubt. Therefore, we can focus on informal explanation.  However, \Coq{} proposes a readable,
rigorous, and computer checked syntax, the \emph{vernacular}, for definitions, lemmas and theorems and when we provide definitions in this paper, they are associated with
expressions stated in the vernacular provide in the appendix.  The vernacular should be seen as a XXI$^{st}$ century version of Leibniz' \emph{characterica universalis} or Frege's
\emph{Begriffsshrift}~\cite{frege67:_from_frege_to}.

\subsubsection{Decomposing an object}

The principle of coinduction is based on the greatest fixpoint of the definition, that is a \emph{coinduction defines a greatest fixpoint} (see~\cite{BertotCasterant04}).  There are
two challenges when one works with such a principle: the difficulty of decomposing infinite objects and the invocation of coinduction.  They are both presented in detail
in~\cite{BertotCasterant04}, but let us describe them in few words. For the first problem, suppose one has a strategy $s$, which is not a leaf; one knows that $s$ is of the form
${\og{a,c, s_l,s_r}\fg}$ for some agent $a$, some choice $c$ and some strategies $s_l$ and $s_r$.  To obtain such a presentation, one uses a mechanism which consists in defining a
function identity on strategy which is a ``clone'' of $fun\ s "=>" s\ end$:\pagebreak[3]

\medskip
\noindent
\coqdockw{Definition} \coqdocid{Strategy\_identity} (\coqdocid{s}:\coqdocid{Strategy}): \coqdocid{Strategy} :=\coqdoceol
\noindent
\coqdocid{match} \coqdocid{s} \coqdocid{with}\coqdoceol
\noindent
| $\og$\coqdocid{f}$\fg$ \ensuremath{\Rightarrow}  $\og$\coqdocid{f}$\fg$\coqdoceol
\noindent
| $\og$\coqdocid{a},\coqdocid{c},\coqdocid{sl},\coqdocid{sr}$\fg$ \ensuremath{\Rightarrow} $\og$\coqdocid{a},\coqdocid{c},\coqdocid{sl},\coqdocid{sr}$\fg$\coqdoceol
\noindent
\coqdocid{end}.\coqdoceol

\medskip

\noindent In other words, the \emph{strategy identity} function is, computationally speaking, the function which associates $\og$\coqdocid{f}$\fg$ with $\og$\coqdocid{f}$\fg$ and
$\og$\coqdocid{a},\coqdocid{c},\coqdocid{sl},\coqdocid{sr}$\fg$ with $\og$\coqdocid{a},\coqdocid{c},\coqdocid{sl},\coqdocid{sr}$\fg$ and not the function which associates $s$ with $s$.
We can prove the \emph{strategy decomposition} lemma:

\medskip
\noindent
\coqdockw{Lemma} \coqdocid{Strategy\_decomposition}: \ensuremath{\forall} \coqdocid{s}: \coqdocid{Strategy},\coqdoceol
\coqdocindent{1.00em}
\coqdocid{Strategy\_identity} \coqdocid{s} = \coqdocid{s}.\coqdoceol

\medskip

Thus when one wants to decompose a strategy $s$, one replaces $s$ by \coqdocid{Strategy\_identity} $s$ and one simplifies the expression, and one gets ${\og a,c, s_l,s_r\fg}$ for
some $a$, $c$, $s_l$ and $s_r$.

\subsubsection{ Invoking coinduction}

The \emph{principle of coinduction} is based on a \emph{tactic}\footnote{A \emph{tactic} is a tool in \Coq{} used to build proofs without using the most elementary constructions.}
called \emph{cofix}.  It consists in assuming the proposition one wants to proof, provided one applies it only on strict sub-objects. In the current implementation of \Coq{}, the
user has to ensure that he invokes it on ``strict'' sub-objects.  This is not always completely trivial and requires a good methodology.  However the \emph{proof checker} (a piece
of software which accepts only correct proofs) verifies that this constraint is fulfilled at the time of checking the proof.

\section{Related works}
\label{sec:rel_works}

This work started after this of Vestergaard~\cite{vestergaard06:IPL} on finite games and finite strategies.  We first developed proofs on finite strategies, but unlike Vestergaard
who based his formalization on fixpoint definitions of predicates, we used only inductive definitions of predicates. Like Vestergaard, we were able to prove the main lemma of
finite extensive games, namely that \emph{backward induction strategy profiles are Nash equilibria}; the script is available at
\url{http://perso.ens-lyon.fr/pierre.lescanne/COQ/INFGAMES/SCRIPTS/finite_games.v}.  Overall, this ``induction based'' presentation allowed us to switch more easily to coinduction
on infinite games.  Beside this, a development in \Coq{} of finite games with an arbitrary number of choices at any node has been made by Le~Roux~\cite{LeRouxPhD08} (p.~83 and
following) and an exploration of common knowledge, induction and Aumann's theorem on rationality has been proposed by Vestergaard, Ono and the
author~\cite{vestergaard06:_lescan_ono}. In~\cite{lescanne07:_mechan_coq}, there is a presentation of a somewhat connected development in \Coq, namely this of the \emph{logic of
  common knowledge}.

Since we are talking about some computational aspects of games, people may make some analogies with other works, let us state what extensive games are not.
\begin{itemize}
\item Extensive games are not \emph{semantic games} as presented in~\cite{s.abramsky94:_games_and_full_compl_for,benthem06:_logic_in_games,lorenz78:_dialog_logik,Locus_Solum_966909}.

\item Extensive games are not \emph{logical games} used in proving properties of automata and protocols~\cite{DBLP:conf/tphol/Merz00,DBLP:conf/provsec/AffeldtTM07}.
\item This work has only loose connection with \emph{algorithmic game theory}~\cite{1296179,daskalakis09:_compl_of_comput_nash_equil}, which is more interested by the complexity of
  the algorithms, especially those which compute equilibria, than by their correction, and does not deal with infinite games.
\end{itemize}

Three published examples of \Coq{} developments with coinduction are this on \emph{real numbers} by Bertot~\cite{bertot07:_affin_funct_and_series_with}, this on \emph{temporal
  logic} by~Coupet-Grimal~\cite{coupet-grimal03:_axiom_of_linear_temp_logic} and this on \emph{hardware verification} by Coupet-Grimal and
Jakubiec~\cite{DBLP:journals/fac/Coupet-GrimalJ04}.  The book~\cite{dowek07:_les_metam_du_calcul} by Dowek gives a
philosophical perspective of using a proof assistant based on type theory in mathematics.

\section*{Acknowledgments}
\label{ack}

I would like to thanks many persons for advice in the course of this research, among them I am especially indebted to Rene Vestergaard, Stephane Le~Roux, Franck Delaplace, Herbert
Gintis, Ralph Matthes, Philippe Audebaud, Olivier Laurent and Alex\-andre Miquel and last but not least Yves Bertot and Pierre Cast\'eran for their beautiful \emph{Coq'Art}.

\section{Conclusion}
\label{sec:concl}

Thanks to coinduction, we have reconciled human reasoning with rational reasoning in infinite extensive games.  In other words, we claim that human agents reason actually by
coinduction when faced to infinite games and are rational.  Moreover we have shown once more the threshold between finiteness and infiniteness and that reasoning on infinite objects
is not the limit when the size goes to infinity of reasoning on finite objects.


\begin{thebibliography}{10}

\bibitem{s.abramsky94:_games_and_full_compl_for}
S.~Abramsky and R.~Jagadeesan.
\newblock Games and full completeness for multiplicative linear logic.
\newblock {\em Journal of Symbolic Logic}, 59:543--574, 1994.

\bibitem{DBLP:conf/provsec/AffeldtTM07}
Reynald Affeldt, Miki Tanaka, and Nicolas Marti.
\newblock Formal proof of provable security by game-playing in a proof
  assistant.
\newblock In Willy Susilo, Joseph~K. Liu, and Yi~Mu, editors, {\em ProvSec},
  volume 4784 of {\em Lecture Notes in Computer Science}, pages 151--168.
  Springer, 2007.

\bibitem{bertot07:_affin_funct_and_series_with}
Yves Bertot.
\newblock Affine functions and series with co-inductive real numbers.
\newblock {\em Mathematical Structures in Computer Science}, 17(nn):37--63,
  2007.

\bibitem{BertotCasterant04}
Yves Bertot and Pierre Cast\'{e}ran.
\newblock {\em Interactive Theorem Proving and Program Development Coq'Art: The
  Calculus of Inductive Constructions}.
\newblock Springer-Verlag, 2004.

\bibitem{Colman_rat_back_ind}
A.~M. Colman.
\newblock {\em Rational Models of Cognition}, chapter Rationality assumptions
  of game theory and the backward induction paradox, pages 353--371.
\newblock Oxford U. Press, 1998.
\newblock Edited by Mike Oaksford and Nick Chater.

\bibitem{colman99:_game_theor_and_its_applic}
Andrew~M. Colman.
\newblock {\em Game theory and its applications in the social and biological
  sciences}.
\newblock London New York : Routledge, 1999.
\newblock Second edition.

\bibitem{DBLP:conf/types/Coquand93}
Thierry Coquand.
\newblock Infinite objects in type theory.
\newblock In Henk Barendregt and Tobias Nipkow, editors, {\em TYPES}, volume
  806 of {\em Lecture Notes in Computer Science}, pages 62--78. Springer, 1993.

\bibitem{coupet-grimal03:_axiom_of_linear_temp_logic}
Solange Coupet-Grimal.
\newblock An axiomatization of linear temporal logic in the calculus of
  inductive constructions.
\newblock {\em J Logic Computation}, 13(6):801--813, 2003.

\bibitem{DBLP:journals/fac/Coupet-GrimalJ04}
Solange Coupet-Grimal and Line Jakubiec.
\newblock Certifying circuits in type theory.
\newblock {\em Formal Asp. Comput.}, 16(4):352--373, 2004.

\bibitem{daskalakis09:_compl_of_comput_nash_equil}
Contantinos Daskalakis, Paul~W. Golberg, and Christos~H. Papadimitriou.
\newblock The complexity of computing a {Nash} equilibrium.
\newblock {\em Communications of the ACM}, 52(2):89--97, 2009.

\bibitem{dowek07:_les_metam_du_calcul}
G.~Dowek.
\newblock {\em Les m\'etamorphoses du calcul}.
\newblock Le Pommier, 2007.
\newblock Grand Prix of Philosophy of the French Academy.

\bibitem{EF-finite-mt}
H.~D. Ebbinghaus and J.~Flum.
\newblock {\em Finite Model Theory}.
\newblock Springer-Verlag, New York, 1995.

\bibitem{fagin93:_finit_model_theor_person_persp}
R.~Fagin.
\newblock Finite model theory -- a personal perspective.
\newblock {\em Theoretical Computer Science}, 116:3--31, 1993.

\bibitem{frege67:_from_frege_to}
Gottlob Frege.
\newblock {\em From Frege to Gödel}, chapter Concept Script.
\newblock Harvard Uni. Press., 1967.

\bibitem{gintis00:_game_theor_evolv}
Herbert Gintis.
\newblock {\em Game Theory Evolving: A Problem-Centered Introduction to
  Modeling Strategic Interaction}.
\newblock Princeton University Press, 2000.

\bibitem{Locus_Solum_966909}
Jean-Yves Girard.
\newblock Locus solum: From the rules of logic to the logic of rules.
\newblock {\em Mathematical. Structures in Comp. Sci.}, 11(3):301--506, 2001.

\bibitem{Kuhn:ExtGamesInfo53}
Harold~W. Kuhn.
\newblock Extensive games and the problem of information.
\newblock {\em Contributions to the Theory of Games {II}}, 1953.
\newblock Reprinted in \cite{Kuhn:97}.

\bibitem{Kuhn:97}
Harold~W. Kuhn, editor.
\newblock {\em Classics in Game Theory}.
\newblock Princeton Uni. Press, 1997.

\bibitem{leininger89:_escal_and_cooop_in_confl_situat}
W.~Leininger.
\newblock Escalation and coooperation in conflict situations.
\newblock {\em J. of Conflict Resolution}, 33:231--254, 1989.

\bibitem{lescanne07:_mechan_coq}
P.~Lescanne.
\newblock Mechanizing epistemic logic with {Coq}.
\newblock {\em Annals of Mathematics and Artificial Intelligence}, 48:15--43,
  2007.

\bibitem{lorenz78:_dialog_logik}
K.~Lorenz and P.~Lorenzen.
\newblock {\em Dialogische Logik}.
\newblock Darmstadt, 1978.

\bibitem{Coq:manual}
\mbox{The Coq development team}.
\newblock {\em The Coq proof assistant reference manual}.
\newblock LogiCal Project, 2007.
\newblock Version 8.1.

\bibitem{DBLP:conf/tphol/Merz00}
Stephan Merz.
\newblock Weak alternating automata in {Isabelle/HOL}.
\newblock In Mark Aagaard and John Harrison, editors, {\em TPHOLs}, volume 1869
  of {\em Lecture Notes in Computer Science}, pages 424--441. Springer, 2000.

\bibitem{Milner89}
R.~Milner.
\newblock {\em Communication and concurrency}.
\newblock Prentice-Hall International Series in Computer Science. Prentice
  Hall, Inc., 1989.

\bibitem{wiki:MCG}
Moli\`ere.
\newblock The middle-class gentleman --- wikisource{,} the free library, 2008.
\newblock [Online; accessed 19-March-2009].

\bibitem{Nipkow-Paulson-Wenzel:2002}
Tobias Nipkow, Lawrence~C. Paulson, and Markus Wenzel.
\newblock {\em Isabelle/HOL --- A Proof Assistant for Higher-Order Logic},
  volume 2283 of {\em LNCS}.
\newblock Springer, 2002.

\bibitem{1296179}
Noam Nisan, Tim Roughgarden, Eva Tardos, and Vijay~V. Vazirani.
\newblock {\em Algorithmic Game Theory}.
\newblock Cambridge University Press, New York, NY, USA, 2007.

\bibitem{oneill86:_inten_escal_and_dollar_auction}
B.~O'Neill.
\newblock International escalation and the dollar auction.
\newblock {\em J. of Conflict Resolution}, 30(33-50), 1986.

\bibitem{osborne94:_cours_game_theory}
M.~J. Osborne and A.~Rubinstein.
\newblock {\em A Course in Game Theory}.
\newblock The MIT Press, Cambridge, Massachusetts, 1994.

\bibitem{osborne04a}
Martin~J. Osborne.
\newblock {\em An Introduction to Game Theory}.
\newblock Oxford, 2004.

\bibitem{cade92-pvs}
S.~Owre, J.~M. Rushby, and N.~Shankar.
\newblock {PVS:} {A} prototype verification system.
\newblock In Deepak Kapur, editor, {\em 11th International Conference on
  Automated Deduction (CADE)}, volume 607 of {\em Lecture Notes in Artificial
  Intelligence}, pages 748--752, Saratoga, {NY}, jun 1992. Springer-Verlag.

\bibitem{DBLP:conf/tcs/Park81}
David Park.
\newblock Concurrency and automata on infinite sequences.
\newblock In Peter Deussen, editor, {\em Theoretical Computer Science}, volume
  104 of {\em Lecture Notes in Computer Science}, pages 167--183. Springer,
  1981.

\bibitem{rosenthal81:_games_of_perfec_infor_predat}
R.~W. Rosenthal.
\newblock Games of perfect information, predatory pricing and the chain-store
  paradox.
\newblock {\em Journal of Economy Theory}, 25:92--100, 1981.

\bibitem{LeRouxPhD08}
St\'ephane~Le Roux.
\newblock {\em Abstraction and Formalization in Game Theory}.
\newblock PhD thesis, \'Ecole normale sup\'erieure de Lyon (France), January
  2008.

\bibitem{Shubik:1971}
Martin Shubik.
\newblock The dollar auction game: A paradox in noncooperative behavior and
  escalation.
\newblock {\em Journal of Conflict Resolution}, 15(1):109--111, 1971.

\bibitem{shustek08:_inter_donal_knuth}
Len Shustek.
\newblock Interview {Donald Knuth}: A life's work interrupted.
\newblock {\em Communications of the ACM}, 51(8):31--37, August 2008.

\bibitem{benthem06:_logic_in_games}
J.~van Benthem.
\newblock {\em Logic in Games}.
\newblock Elsevier, 2006.

\bibitem{vestergaard06:IPL}
Ren{\'e} Vestergaard.
\newblock A constructive approach to sequential {Nash} equilibria.
\newblock {\em Inf. Process. Lett.}, 97:46--51, 2006.

\bibitem{vestergaard06:_lescan_ono}
Rene Vestergaard, Pierre Lescanne, and Hiroakira Ono.
\newblock Pierre lescanne and hiroakira ono.
\newblock Research Report JAIST/IS-RR-2006-009, Japan InsT. of Science and
  Technology, 2006.

\end{thebibliography}

\appendix

\section{Equalities}
\label{sec:eq}

\emph{Leibniz equality} says that $x=y$ if and only if, for every predicate $P$, $P(x)$ implies $P(y)$.  \emph{Extensional equality} says that $f=g$ if and only if, for all $x$,
$f(x)=g(x)$. In general, knowing a (recursive) definition of $f$ and a (recursive) definition of~$g$ is not enough to decide whether $f=g$ or $f\neq g$.  For instance, no one knows
how to prove that the two functions:
\begin{eqnarray*}
  f(1)&=&1\\
  f(2x) &=& f(x)\\
  f(2x+1) &=& f(3x+2).
\end{eqnarray*}
and
\[g(x)~=~1\] are equal, despite it is more likely that they are.  More generally, there is no algorithm (no rigorous reasoning) which decides whether a given function $h$ is equal
to the above function $g$ or not.  Thus \emph{extensional equality} is not decidable.  Saying that two sequences that have equal elements are equal requires \emph{extensional
  equality} and it makes sense to reject such an equality when reasoning finitely about infinite objects, like human agents would do.

\section{Some definition and proposition in \Coq{} vernacular} \label{sec:coq_vern}

\subsubsection*{The notation of functions in the \Coq{} vernacular.}

In traditional mathematics, the result of applying a function $f$ to the value $x$ is written $f(x)$ and the result of applying $f$ to $x$ and $y$ is written $f(x,y)$, this can be
considered as the result of applying $f$ to $x$ then to $y$ and written $f(x)(y)$.  In the \Coq{} vernacular, as in type theory, one writes $f~x$ instead of $f(x)$ and $f~x~y$
instead $f(x)(y)$ or instead of $f(x,y)$ and $f~x~y~z$ instead $f(x)(y)(z)$ or $f(x,y,z)$, because this saves parentheses and commas and because functions are everywhere as the
core of the formalization.  But after all, this is not deep, just a matter of style and \Coq{} accepts syntactic shorthands to avoid these notations when others are desirable.

\subsubsection*{The definitions}
\label{sec:def}

The following definition are in the \Coq{} vernacular.

The definition of \coqdocid{History} is 

\medskip

\noindent
\coqdockw{CoInductive} \coqdocid{History}: \coqdocid{Set} :=\coqdoceol
\noindent
| \coqdocid{HNil}: \coqdocid{History}\coqdoceol
\noindent
| \coqdocid{HCons}: \coqdocid{Choice} \ensuremath{\rightarrow} \coqdocid{History} \ensuremath{\rightarrow} \coqdocid{History}.\coqdoceol

\medskip 

The definition of \emph{bisimilarity} in the \Coq{}is (Notice
the use of the notation (\coqdocid{HCons} \coqdocid{a} \coqdocid{h}) for \coqdocid{HCons}(\coqdocid{a})(\coqdocid{h})):

\medskip
\noindent
\coqdockw{CoInductive} \coqdocid{hBisimilar}: \coqdocid{History}  \ensuremath{\rightarrow}\coqdocid{History} \ensuremath{\rightarrow} \coqdocid{Prop} :=\coqdoceol
\noindent
| \coqdocid{bisim\_HNil}: \coqdocid{hBisimilar} \coqdocid{HNil} \coqdocid{HNil}\coqdoceol
\noindent
| \coqdocid{bisim\_HCons}: \ensuremath{\forall} (\coqdocid{a}:\coqdocid{Choice})(\coqdocid{h} \coqdocid{h'}:\coqdocid{History}),\coqdoceol
\coqdocindent{1.00em}
\coqdocid{hBisimilar} \coqdocid{h} \coqdocid{h'} \ensuremath{\rightarrow} \coqdocid{hBisimilar} (\coqdocid{HCons} \coqdocid{a} \coqdocid{h}) (\coqdocid{HCons} \coqdocid{a} \coqdocid{h'}).\coqdoceol

\medskip

The definition of \emph{Game} is:

\medskip

\noindent
\coqdockw{CoInductive} \coqdocid{Game} : \coqdocid{Set} :=\coqdoceol
\noindent
| \coqdocid{gLeaf}:  \coqdocid{Utility\_fun} \ensuremath{\rightarrow} \coqdocid{Game}\coqdoceol
\noindent
| \coqdocid{gNode} : \coqdocid{Agent} \ensuremath{\rightarrow} \coqdocid{Game} \ensuremath{\rightarrow} \coqdocid{Game} \ensuremath{\rightarrow} \coqdocid{Game}.\coqdoceol

\medskip

The definition of \emph{Strategy} is:

\medskip
\noindent
\coqdockw{CoInductive}
 \coqdocid{Strategy} : \coqdocid{Set} :=\coqdoceol
\noindent
| \coqdocid{sLeaf} : \coqdocid{Utility\_fun} \ensuremath{\rightarrow} \coqdocid{Strategy}\coqdoceol
\noindent
| \coqdocid{sNode} : \coqdocid{Agent} \ensuremath{\rightarrow} \coqdocid{Choice} \ensuremath{\rightarrow} \coqdocid{Strategy} \ensuremath{\rightarrow} \coqdocid{Strategy} \ensuremath{\rightarrow} \coqdocid{Strategy}.\coqdoceol

\medskip

The function \emph{s2u} is defined as 

\medskip

\noindent
\coqdockw{CoInductive} \coqdocid{s2u} : \coqdocid{Strategy} \ensuremath{\rightarrow} \coqdocid{Agent} \ensuremath{\rightarrow} \coqdocid{Utility} \ensuremath{\rightarrow} \coqdocid{Prop} :=\coqdoceol
\noindent
| \coqdocid{s2uLeaf}: \ensuremath{\forall} \coqdocid{a} \coqdocid{f}, \coqdocid{s2u} ($\og$ \coqdocid{f}$\fg$) \coqdocid{a} (\coqdocid{f} \coqdocid{a})\coqdoceol
\noindent
| \coqdocid{s2uLeft}: \ensuremath{\forall}  (\coqdocid{a} \coqdocid{a'}:\coqdocid{Agent}) (\coqdocid{u}:\coqdocid{Utility}) (\coqdocid{sl} \coqdocid{sr}:\coqdocid{Strategy}),\coqdoceol
\coqdocindent{2.00em}
\coqdocid{s2u} \coqdocid{sl} \coqdocid{a} \coqdocid{u}  \ensuremath{\rightarrow} \coqdocid{s2u} ($\og$ \coqdocid{a'},\coqdocid{l},\coqdocid{sl},\coqdocid{sr}$\fg$) \coqdocid{a} \coqdocid{u}  \coqdoceol
\noindent
| \coqdocid{s2uRight}: \ensuremath{\forall} (\coqdocid{a} \coqdocid{a'}:\coqdocid{Agent}) (\coqdocid{u}:\coqdocid{Utility}) (\coqdocid{sl} \coqdocid{sr}:\coqdocid{Strategy}),\coqdoceol
\coqdocindent{2.00em}
\coqdocid{s2u} \coqdocid{sr} \coqdocid{a} \coqdocid{u} \ensuremath{\rightarrow} \coqdocid{s2u} ($\og$ \coqdocid{a'},\coqdocid{r},\coqdocid{sl},\coqdocid{sr}$\fg$) \coqdocid{a} \coqdocid{u}.\coqdoceol

\medskip

\emph{LeadstoLeaf} is an \textbf{inductive}.

\medskip
\noindent
\coqdockw{Inductive} \coqdocid{LeadsToLeaf}: \coqdocid{Strategy} \ensuremath{\rightarrow} \coqdocid{Prop} :=\coqdoceol
\noindent
| \coqdocid{LtLLeaf}: \ensuremath{\forall} \coqdocid{f}, \coqdocid{LeadsToLeaf} ($\og$ \coqdocid{f}$\fg$)\coqdoceol
\noindent
| \coqdocid{LtLLeft}: \ensuremath{\forall} (\coqdocid{a}:\coqdocid{Agent})(\coqdocid{sl}: \coqdocid{Strategy}) (\coqdocid{sr}:\coqdocid{Strategy}),\coqdoceol
\coqdocindent{2.00em}
\coqdocid{LeadsToLeaf} \coqdocid{sl} \ensuremath{\rightarrow} \coqdocid{LeadsToLeaf} ($\og$ \coqdocid{a},\coqdocid{l},\coqdocid{sl},\coqdocid{sr}$\fg$)\coqdoceol
\noindent
| \coqdocid{LtLRight}: \ensuremath{\forall} (\coqdocid{a}:\coqdocid{Agent})(\coqdocid{sl}: \coqdocid{Strategy}) (\coqdocid{sr}:\coqdocid{Strategy}),\coqdoceol
\coqdocindent{2.50em}
\coqdocid{LeadsToLeaf} \coqdocid{sr} \ensuremath{\rightarrow} \coqdocid{LeadsToLeaf} ($\og$ \coqdocid{a},\coqdocid{r},\coqdocid{sl},\coqdocid{sr}$\fg$).\coqdoceol

\medskip 

Two lemmas about \emph{LeadsToLeaf} and $s2u$:

\begin{itemize}
\item[$`(?)$] $`A a\ s, LeadsToLeaf\ s "->" `E u: Utility, s2u\ s\ a\ u $.
\item[$`(?)$] $`A a\ u\ v\ s, LeadsToLeaf\ s "->" s2u\ s\ a\ u "->" s2u\ s\ a\ v "->" u=v$.
\end{itemize}

The predicate which we called \altl{} is formally defined in the vernacular as follows:

\medskip
\noindent
\coqdockw{CoInductive} \coqdocid{AlwLeadsToLeaf}: \coqdocid{Strategy} \ensuremath{\rightarrow} \coqdocid{Prop} :=\coqdoceol
\noindent
| \coqdocid{ALtLeaf} : \ensuremath{\forall} (\coqdocid{f}:\coqdocid{Utility\_fun}), \coqdocid{AlwLeadsToLeaf} ($\og$\coqdocid{f}$\fg$)\coqdoceol
\noindent
| \coqdocid{ALtL} : \ensuremath{\forall} (\coqdocid{a}:\coqdocid{Agent})(\coqdocid{c}:\coqdocid{Choice})(\coqdocid{sl} \coqdocid{sr}:\coqdocid{Strategy}),\coqdoceol
\coqdocindent{2.00em}
\coqdocid{LeadsToLeaf} ($\og$\coqdocid{a},\coqdocid{c},\coqdocid{sl},\coqdocid{sr}$\fg$) \ensuremath{\rightarrow} \coqdocid{AlwLeadsToLeaf} \coqdocid{sl} \ensuremath{\rightarrow}\coqdocid{AlwLeadsToLeaf} \coqdocid{sr} \ensuremath{\rightarrow} \coqdoceol
\coqdocindent{2.00em}
\coqdocid{AlwLeadsToLeaf} ($\og$\coqdocid{a},\coqdocid{c},\coqdocid{sl},\coqdocid{sr}$\fg$).\coqdoceol

\medskip
Bisimilarity of \emph{strategies} is defined \textbf{coinductively}:
\medskip

\noindent
\coqdockw{CoInductive} \coqdocid{sBisimilar}: \coqdocid{Strategy} \ensuremath{\rightarrow} \coqdocid{Strategy} \ensuremath{\rightarrow} \coqdocid{Prop} :=\coqdoceol
\noindent
| \coqdocid{bisim\_sLeaf}: \ensuremath{\forall} \coqdocid{f}, \coqdocid{sBisimilar} ($\og$\coqdocid{f}$\fg$) ($\og$\coqdocid{f}$\fg$)\coqdoceol
\noindent
| \coqdocid{bisim\_sNode}: \ensuremath{\forall} \coqdocid{a} \coqdocid{c} \coqdocid{sl} \coqdocid{sl'} \coqdocid{sr} \coqdocid{sr'},\coqdoceol
\coqdocindent{1.00em}
\coqdocid{sBisimilar} \coqdocid{sl} \coqdocid{sl'} \ensuremath{\rightarrow} \coqdocid{sBisimilar} \coqdocid{sr} \coqdocid{sr'} \ensuremath{\rightarrow}  \coqdocid{sBisimilar} ($\og$\coqdocid{a},\coqdocid{c},\coqdocid{sl},\coqdocid{sr}$\fg$) ($\og$\coqdocid{a}, \coqdocid{c},\coqdocid{sl'},\coqdocid{sr'}$\fg$).\coqdoceol

\medskip

\emph{SGPE} is defined \textbf{coinductively} as follows:

  \medskip

  \noindent \coqdockw{CoInductive} \coqdocid{SGPE}: \coqdocid{Strategy} \ensuremath{\rightarrow} \coqdocid{Prop} :=\coqdoceol
  \noindent | \coqdocid{SGPE\_leaf}: \ensuremath{\forall} \coqdocid{f}:\coqdocid{Utility\_fun}, \coqdocid{SGPE} ($\og$\coqdocid{f}$\fg$)\coqdoceol
  \noindent | \coqdocid{SGPE\_left}: \ensuremath{\forall} (\coqdocid{a}:\coqdocid{Agent})(\coqdocid{u} \coqdocid{v}: \coqdocid{Utility}) (\coqdocid{sl} \coqdocid{sr}:
  \coqdocid{Strategy}), \coqdoceol \coqdocindent{2.00em} \coqdocid{AlwLeadsToLeaf} ($\og$\coqdocid{a},\coqdocid{l},\coqdocid{sl},\coqdocid{sr}$\fg$) \ensuremath{\rightarrow}
  \coqdoceol \coqdocindent{2.00em} \coqdocid{SGPE} \coqdocid{sl} \ensuremath{\rightarrow} \coqdocid{SGPE} \coqdocid{sr} \ensuremath{\rightarrow} \coqdoceol \coqdocindent{2.00em}
  \coqdocid{s2u} \coqdocid{sl} \coqdocid{a} \coqdocid{u} \ensuremath{\rightarrow} \coqdocid{s2u} \coqdocid{sr} \coqdocid{a} \coqdocid{v} \ensuremath{\rightarrow} (\coqdocid{v}
  \leut \coqdocid{u}) \ensuremath{\rightarrow} \coqdoceol \coqdocindent{2.00em} \coqdocid{SGPE} ($\og$\coqdocid{a},\coqdocid{l},\coqdocid{sl},\coqdocid{sr}$\fg$)\coqdoceol
  \noindent | \coqdocid{SGPE\_right}: \ensuremath{\forall} (\coqdocid{a}:\coqdocid{Agent}) (\coqdocid{u} \coqdocid{v}:\coqdocid{Utility}) (\coqdocid{sl} \coqdocid{sr}:
  \coqdocid{Strategy}), \coqdoceol \coqdocindent{2.00em} \coqdocid{AlwLeadsToLeaf} ($\og$\coqdocid{a},\coqdocid{r},\coqdocid{sl},\coqdocid{sr}$\fg$) \ensuremath{\rightarrow}
  \coqdoceol \coqdocindent{2.00em} \coqdocid{SGPE} \coqdocid{sl} \ensuremath{\rightarrow} \coqdocid{SGPE} \coqdocid{sr} \ensuremath{\rightarrow} \coqdoceol \coqdocindent{2.00em}
  \coqdocid{s2u} \coqdocid{sl} \coqdocid{a} \coqdocid{u} \ensuremath{\rightarrow} \coqdocid{s2u} \coqdocid{sr} \coqdocid{a} \coqdocid{v} \ensuremath{\rightarrow} (\coqdocid{u}
  \leut \coqdocid{v}) \ensuremath{\rightarrow} \coqdoceol \coqdocindent{2.00em} \coqdocid{SGPE} ($\og$\coqdocid{a},\coqdocid{r},\coqdocid{sl},\coqdocid{sr}$\fg$). \coqdoceol

\medskip

\noindent
\coqdockw{Lemma} \coqdocid{AlwLeadsToLeaf\_Preservation}: \ensuremath{\forall} (\coqdocid{a}:\coqdocid{Agent})(\coqdocid{s} \coqdocid{s'}:\coqdocid{Strategy}),\coqdoceol
\coqdocindent{1.00em}
\coqdocid{AlwLeadsToLeaf} \coqdocid{s} \ensuremath{\rightarrow} \coqdocid{s} \conva \coqdocid{s'} \ensuremath{\rightarrow} \coqdocid{AlwLeadsToLeaf} \coqdocid{s'}.\coqdoceol

\medskip

\medskip

\noindent
\coqdockw{Definition} \coqdocid{NashEq} (\coqdocid{s}: \coqdocid{Strategy}): \coqdocid{Prop} := \coqdoceol
\coqdocindent{1.00em}
\ensuremath{\forall} \coqdocid{a} \coqdocid{s'} \coqdocid{u} \coqdocid{u'}, \coqdocid{s'}\conva\coqdocid{s} \ensuremath{\rightarrow} \coqdoceol
\coqdocindent{2.00em}
\coqdocid{LeadsToLeaf} \coqdocid{s'} \ensuremath{\rightarrow} (\coqdocid{s2u} \coqdocid{s'} \coqdocid{a} \coqdocid{u'}) \ensuremath{\rightarrow} \coqdoceol
\coqdocindent{2.00em}
\coqdocid{LeadsToLeaf} \coqdocid{s} \ensuremath{\rightarrow} (\coqdocid{s2u} \coqdocid{s} \coqdocid{a} \coqdocid{u}) \ensuremath{\rightarrow} \coqdoceol
\coqdocindent{2.00em}
(\coqdocid{u'} \leut \coqdocid{u}).\coqdoceol

\medskip




\end{document}